\documentclass[10pt,conference,anonymous]{IEEEtran}
\IEEEoverridecommandlockouts
\newcommand{\RQone}{How effectively does \df reduce raw sanitizer reports to confirmed commit-induced bugs?}
\newcommand{\RQtwo}{What runtime cost and code coverage does \df achieve?}
\newcommand{\RQthree}{What types of commit-induced memory-safety bugs does \df detect?}
\newcommand{\RQfour}{Why does \df produce false positives?}

\usepackage{pbalance}
\usepackage{cite}
\usepackage{amsmath,amssymb,amsfonts}
\usepackage{graphicx}
\usepackage{textcomp}
\usepackage{xcolor}
\usepackage{booktabs} % For modern, less cluttered tables
\usepackage{orcidlink}
\usepackage[]{caption}
\usepackage{subcaption}
\usepackage{multirow}
\usepackage{tabularx}
\usepackage{url}
\usepackage{balance}
\usepackage[font=small]{caption}
\usepackage{comment}
\usepackage{makecell}

\usepackage{comment}
\usepackage{makecell}

\usepackage{tikz}
\usepackage{pgfplots}
\pgfplotsset{compat=1.18}
\usepgfplotslibrary{statistics}
\usepackage{enumitem}
\usepackage{fancyhdr}
\usepackage{algorithm}
\usepackage{algpseudocode}
\usepackage{threeparttable}
\usepackage{xspace}
\usepackage{soul}
\usepackage{color,xcolor}
\usepackage{listings}
\usepackage[most]{tcolorbox}
\usepackage{hyperref}
\usepackage{pgf-pie}

\newcommand{\oldfs}{\textsc{FuzzSlice}\xspace}

\newcommand{\df}{{\textsc{{CommitGuard}}}\xspace}

\tcbset{
  stepbox/.style={
    colback=gray!20,
    colframe=gray,
    boxrule=0pt,
    arc=0mm,
    left=1mm, right=1mm, top=0.5mm, bottom=0.5mm,
    fontupper=\bfseries,
    enhanced,
    sharp corners,
  },
  myrqstyle/.style={
    enhanced,
    colback=gray!10,
    colframe=gray!30,
    boxrule=0.4pt,
    arc=1mm,
    left=4pt,
    right=4pt,
    top=2pt,
    bottom=2pt,
    boxsep=4pt,
    fontupper=\bfseries\scshape,
    before skip=6pt,
    after skip=4pt,
  }
}

\lstdefinestyle{mystyle}{
  backgroundcolor=\color{backcolour},
  commentstyle=\color{codegreen},
  keywordstyle=\color{magenta},
  numberstyle=\tiny\color{codegray},
  stringstyle=\color{codepurple},
  basicstyle=\ttfamily\footnotesize,
  breakatwhitespace=false,
  breaklines=false,
  captionpos=b,
  keepspaces=true,
  numbers=left,
  numbersep=5pt,
  showspaces=false,
  showstringspaces=false,
  showtabs=false,
  tabsize=2,
  frame=single,
  rulecolor=\color{bordercolour},
  framerule=0.8pt,
  columns=fullflexible,
  xleftmargin=1.5em,
  framexleftmargin=1.5em
}

\lstdefinestyle{mystyle}{
  backgroundcolor=\color{backcolour},
  commentstyle=\color{codegreen},
  keywordstyle=\color{magenta},
  numberstyle=\tiny\color{codegray},
  stringstyle=\color{codepurple},
  basicstyle=\ttfamily\footnotesize,
  breakatwhitespace=false,
  breaklines=false,
  captionpos=b,
  keepspaces=true,
  numbers=left,
  numbersep=5pt,
  showspaces=false,
  showstringspaces=false,
  showtabs=false,
  tabsize=2,
  frame=single,
  rulecolor=\color{bordercolour},
  framerule=0.8pt,
  columns=fullflexible,
  xleftmargin=1.5em,
  framexleftmargin=1.5em
}

\newtcolorbox{boldheading}{
  colback=white, colframe=black, fontupper=\large, boxrule=1pt
}

\definecolor{codegreen}{rgb}{0,0.6,0}
\definecolor{codegray}{rgb}{0.5,0.5,0.5}
\definecolor{codepurple}{rgb}{0.58,0,0.82}
\definecolor{backcolour}{rgb}{1,1,1} % White background
\definecolor{bordercolour}{rgb}{0.8,0.8,0.8} % Light gray border

\definecolor{diffredbg}{RGB}{255,230,230}
\definecolor{diffgreenbg}{RGB}{230,255,230}

\newcommand{\fulldiffadd}[1]{%
  \rlap{\colorbox{diffgreenbg}{\strut\hspace{\dimexpr\linewidth-2\fboxsep\relax}}}%
  #1%
}
\newcommand{\fulldiffdel}[1]{%
  \rlap{\colorbox{diffredbg}{\strut\hspace{\dimexpr\linewidth-2\fboxsep\relax}}}%
  #1%
}

\def\BibTeX{{\rm B\kern-.05em{\sc i\kern-.025em b}\kern-.08em
    T\kern-.1667em\lower.7ex\hbox{E}\kern-.125emX}}
\begin{document}

% \title{\df: Verifying Code Commits through Differential Slice based Fuzzing}
% \title{\df: Differential Slice-Based Fuzzing for Commit-Level Verification}

\title{\df: Differential Slice Fuzzing for Commit-Induced Bug Detection}
%Author list
% \author{\IEEEauthorblockN{Aniruddhan Murali}
% \IEEEauthorblockA{\textit{Cheriton School of Computer Science} \\
% \textit{University of Waterloo}\\
% Waterloo, Canada\orcidlink{0000-0002-4405-1657}}
% \and
% \IEEEauthorblockN{Noble Saji Mathews}
% \IEEEauthorblockA{\textit{Cheriton School of Computer Science} \\
% \textit{University of Waterloo}\\
% Waterloo, Canada\orcidlink{0000-0003-2266-8848}}
% \and
% \IEEEauthorblockN{Mahmoud Alfadel}
% \IEEEauthorblockA{\textit{Department of Computer Science} \\
% \textit{University of Calgary}\\
% Calgary, Canada\orcidlink{0000-0002-2621-6104}}
% \and
% \IEEEauthorblockN{Meiyappan Nagappan}
% \IEEEauthorblockA{\textit{Cheriton School of Computer Science} \\
% \textit{University of Waterloo}\\
% Waterloo, Canada\orcidlink{0000-0003-4533-4728}}
% \and
% \IEEEauthorblockN{Meng Xu}
% \IEEEauthorblockA{\textit{Cheriton School of Computer Science} \\
% \textit{University of Waterloo}\\
% Waterloo, Canada\orcidlink{0009-0001-6364-4837}}
% }

\author{
  \IEEEauthorblockN{Aniruddhan Murali\orcidlink{0000-0002-4405-1657}}
  \IEEEauthorblockA{\footnotesize \textit{Cheriton School of Computer Science} \\
    \textit{University of Waterloo}\\
    Waterloo, Canada}
  \and
  \IEEEauthorblockN{Noble Saji Mathews\orcidlink{0000-0003-2266-8848}}
  \IEEEauthorblockA{\footnotesize \textit{Cheriton School of Computer Science} \\
    \textit{University of Waterloo}\\
    Waterloo, Canada}
  \and
  \IEEEauthorblockN{Mahmoud Alfadel\orcidlink{0000-0002-2621-6104}}
  \IEEEauthorblockA{\footnotesize \textit{Department of Computer Science} \\
    \textit{University of Calgary}\\
    Calgary, Canada}
  \and
  \IEEEauthorblockN{Meiyappan Nagappan\orcidlink{0000-0003-4533-4728}}
  \IEEEauthorblockA{\footnotesize \textit{Cheriton School of Computer Science} \\
    \textit{University of Waterloo}\\
    Waterloo, Canada}
}
\maketitle

\begin{abstract}
Modern software systems evolve through frequent commits that implement bug fixes, features,  and security patches. Although code review and testing are widely used to check these changes, they often provide limited assurance for memory-safety issues. Code reviewers may miss subtle boundary, lifetime, or initialization errors, while existing tests may not exercise the specific paths affected by a commit. Fuzzing is effective at exposing such bugs, but applying it to every commit remains impractical because whole-program fuzzing is expensive, requires suitable harnesses, and may still fail to reach the code changed by a commit.

In this paper, we introduce \df, a commit-aware differential slice-based fuzzing approach for verifying code changes. 
The key insight behind \df is that the pre-commit version of a modified function can serve as a behavioral baseline for interpreting bugs found after the commit. 
For each target commit, \df identifies modified functions, extracts compilable code slices from both the pre-commit and post-commit versions, and fuzzes the paired slices independently. 
It then compares sanitizer reports across the two versions and reports bugs that emerge \textit{only} in the post-commit version as candidate commit-induced bugs.
We evaluate \df on 300 commits from popular open-source projects. 
Slice fuzzing initially produces 518 sanitizer reports across these commits. By comparing pre-commit and post-commit slices, \df narrows this large output to 7 candidate commit-induced bug reports that require manual triage. Manual validation confirms 5 of these reports as real bugs that were fixed by developers of the examined projects after we reported them, while only 2 reports were classified as false positives. \df analyzes a commit in 32.4 minutes on average and achieves 75.36\% average coverage of modified functions.

Our results show that \df makes commit-level fuzzing practical by turning noisy slice-fuzzing output into a small set of high-confidence commit-induced reports. 
This \textit{on-demand} commit testing framework helps developers focus review effort on risky commits without whole-program fuzzing or manually written tests.
\end{abstract}

\begin{IEEEkeywords}
Fuzzing, Code commits, Differential Analysis, Vulnerability Detection.
\end{IEEEkeywords}

\section{Introduction}
\label{sec:introduction}

Modern software systems evolve through frequent code changes, or commits, that implement bug fixes, new features and security patches~\cite{duvall2007continuous, canfora2011long, marinescu2012assessing, godfrey2002tracking}. This process creates substantial code churn, with large codebases accumulating thousands of commits over time and often modifying complex, tightly coupled components~\cite{hindle2008large, alali2008s}. Ensuring that these commits do not introduce memory-safety bugs is a central challenge in software maintenance. Code review remains the primary mechanism for checking commits, but it is manual, time-consuming, and limited by reviewer expertise and cognitive load~\cite{edmundson2013empirical}. Subtle boundary, lifetime, and initialization errors can evade review, especially when a change appears small or localized~\cite{koru2008investigation}. As a result, commit-induced bugs may persist long after integration, despite passing standard review processes~\cite{canfora2011long}. This gap motivates automated, commit-aware verification techniques that can complement human review and scale to modern development practices.

Existing automated techniques provide useful support, but they do not fully solve commit-level verification. Static analysis tools are widely used to detect potentially harmful bugs early in the software development lifecycle~\cite{Ayewah2008Aug, louridas2006static, zheng2006value, vassallo2020developers}. However, these tools often produce a large number of warnings, many of which are false positives~\cite{Kang2022May, Nadeem2012Mar, Park2016May, Aloraini2017Sep}. This shifts the burden back to developers, who must triage warnings and decide which locations in a commit require urgent attention. Dynamic analysis can provide stronger evidence by checking whether code changes manifest failures at runtime. Existing project tests help, but verifying a new commit often requires additional tests that exercise the changed paths, which is labor-intensive~\cite{Levin2017Sep, Alegroth2016Feb}.

Fuzzing is an attractive dynamic analysis technique for exposing memory-safety bugs, but whole-program fuzzing is poorly suited for commit-level verification. Fuzzing explores program behavior by generating and mutating inputs, and it has been effective at uncovering bugs in real systems~\cite{Li2018Dec, csallner2005check}. However, applying fuzzing to every commit is difficult. It requires suitable harnesses and can consume substantial computational resources. Also, it may hit coverage plateaus before reaching the code affected by a commit~\cite{gao2023beyond, bohme2020fuzzing}. 
Such limitations make it hard to use whole-program fuzzing as a fast feedback mechanism during code review.

Targeted fuzzing techniques reduce this cost, but they usually assume that the target location is already known. Directed fuzzers guide input generation toward specific locations, often to reproduce known bugs or validate static-analysis reports~\cite{bohme2017directed, Chen2018Oct, DAFL}. Distance-based techniques prioritize inputs that reduce the distance to a target, thereby increasing the chance of exercising the relevant code. Slicing-based fuzzers take a different approach. They construct isolated program slices around suspicious locations and test these slices independently~\cite{FuzzSlice, TPSlice, harman1995using, Chen2022Nov}. By starting execution closer to the target code, slicing-based fuzzing increases the likelihood of exercising suspicious behavior. Although such slices do not produce end-to-end inputs for the full program, their relaxed constraints assess whether a reported bug is plausible~\cite{FuzzSlice}.

Commit-level verification presents a different problem from validating a known warning or bug location. A commit is a code change whose behavioral impact is not known in advance. It may modify multiple lines, functions, or modules, and the verification task is not only to expose a failure, but also to determine whether that failure was introduced by the commit. Fuzzing an isolated slice can produce reports that are hard to interpret without full-program context, while fuzzing the whole program for every commit is too costly. A commit-level technique, therefore, needs more than faster fuzzing. It must identify what changed, construct comparable fuzzing targets across versions, and distinguish newly introduced bugs from bugs or artifacts that already existed before the change (i.e., before the commit has been applied).

To address this problem, we introduce \df, a commit-aware differential slice-based fuzzing method for verifying code changes. The key insight behind \df is that the pre-commit version of a modified function can serve as a behavioral baseline for interpreting bugs found after the commit has been applied. 
Given a target commit and a baseline commit from the same repository, such as the immediate predecessor of the target commit, \df identifies modified functions, extracts compilable code slices from both versions, fuzzes the paired slices independently, and compares their sanitizer reports. 
A report is treated as a candidate commit-induced bug \textit{only} when it appears in the post-commit slice but not in the corresponding pre-commit slice. This differential design turns slice fuzzing from a general bug-finding technique into an \textit{on-demand} commit-verification method. It focuses developer attention on bugs likely introduced by recent code changes while preserving the efficiency benefits of slice-based fuzzing.

We clarify that in this paper, \textit{verify} refers to validation via differential fuzz testing. Therefore, verification results provide empirical evidence based on explored inputs and do not constitute a formal guarantee of correctness. In summary, our study makes the following contributions:
\begin{itemize}
\item We formulate commit-level verification as a differential slice-fuzzing problem. 
Instead of fuzzing whole program versions or validating known warning locations, \df constructs paired pre-commit and post-commit slices and uses differences in sanitizer-observed behavior as an oracle for commit-induced bugs.

% \item We present \df, a commit-aware approach that automatically identifies modified functions, extracts comparable slices across versions, fuzzes the paired slices, and reports bugs that appear after the commit.

\item We evaluate \df on 300 commits from open-source projects. \df narrows 518 raw sanitizer reports from slice fuzzing to only 7 reports identified as candidate commit-induced bug reports. Manual validation by developers of the examined projects confirms 5 of these reports as true bugs, which were fixed by developers after being reported.
% , while only 2 reports were classified as false positives.
% ~\cite{openssl-buffer-overflow-fix, libpcap-buffer-overread-fix, leptonica-integer-overflow-fix, leptonica-mem-leak-fix}.

\item We show that \df is efficient enough for commit-level analysis, requiring 32.4 minutes per commit on average while achieving 75.36\% average coverage of modified functions. We provide a qualitative analysis of the confirmed bugs and false positives to explain the merits and limitations of differential slice-based fuzzing.
% \item Our replication package and datasets are available \href{https://osf.io/j5gce/overview?view_only=31a823ecb1ab44b6ba0e16c7e9c9e4ee}{here}.
\end{itemize}

\section{A Motivating Example}
\label{sec}

This section illustrates why commit-level verification benefits from comparing program behavior before and after a code change. A bug observed when fuzzing a changed function may be difficult to interpret in isolation, because the same behavior may have existed before the commit or may arise from the relaxed context of slice execution. \df addresses this problem by fuzzing paired slices from the pre-commit and post-commit versions and reporting only bugs that newly appear after the commit.

To motivate this idea, we use a commit in the \textsc{file} utility of UNIX systems. The commit aims to extend build-ID reporting to support 8-byte IDs generated by \texttt{lld}. In UNIX systems, \textsc{file} identifies a file's type by inspecting its contents rather than relying on its name or extension. The relevant change modifies the \texttt{do\_bid\_note} function in the \texttt{src/readelf.c} file, which parses the \texttt{.notes} section of an ELF binary and reports GNU build-ID metadata.

\begin{lstlisting}[language=C,
caption={Simplified excerpt from a commit in the UNIX file utility that introduced a stack buffer overflow. The changed guard allows values larger than 20 to reach a copy into a fixed 20-byte stack buffer.},
label={lst-motiv},
breaklines=true,
numbers=left,
numberstyle=\tiny,
xleftmargin=2em,
basicstyle=\ttfamily\footnotesize,
commentstyle=\ttfamily\footnotesize,
escapeinside={(@}{@)}
]
do_bid_note(unsigned char* nbuf,
uint32_t type, uint32_t namesz, uint32_t descsz,
size_t noff, size_t doff, int flags) {
if (namesz == 4 &&
strcmp((char )&nbuf[noff], "GNU") == 0 &&
(@\fulldiffdel{- (descsz == 16 || descsz == 20)}@)
(@\fulldiffadd{+ (descsz >= 4 || descsz <= 20)}@)
) {
  uint8_t desc[20];
  ...
  memcpy(desc, &nbuf[doff], descsz);
}
}
\end{lstlisting}

Listing~\ref{lst-motiv} shows the key lines from the commit~\cite{file-cve-2017-1000249}. The function processes an ELF note stored in \texttt{nbuf}. The parameters \texttt{noff} and \texttt{doff} point to the note name and note description, respectively, while \texttt{namesz} and \texttt{descsz} give their sizes.
Lines 4--5 first check that the note is a GNU build-ID note: the name size must be 4 and the name at \texttt{nbuf[noff]} must be \texttt{"GNU"}. The changed condition on lines 6--7 then decides which description sizes are accepted before the build-ID data is copied.

In the pre-commit version, line 6 accepts only build-ID lengths of exactly 16 or 20 bytes. Both values fit within the fixed 20-byte stack buffer \texttt{desc}, declared on line 9. Therefore, when execution reaches the \texttt{memcpy} on line 11, the number of bytes copied into \texttt{desc} is bounded by the buffer size.

The commit changes the guard to \texttt{descsz >= 4 || descsz <= 20} on line 7. The developer's intent appears to be accepting build-ID sizes in the range from 4 to 20 bytes, including the new 8-byte case introduced by the commit. However, the guard uses disjunction instead of conjunction. This makes the condition true for every value of \texttt{descsz}, i.e., values below 4 satisfy \texttt{descsz <= 20}, values above 20 satisfy \texttt{descsz >= 4}, and values between 4 and 20 satisfy both. As a result, an input with \texttt{descsz > 20} can pass the guard.

Once such an input reaches the branch body, the function copies \texttt{descsz} bytes from \texttt{nbuf[doff]} into the fixed 20-byte stack buffer \texttt{desc} on line 11. If \texttt{descsz} is greater than 20, the copy writes past the end of \texttt{desc}, producing a stack buffer overflow. This vulnerability was assigned CVE-2017-1000249~\cite{file-cve-2017-1000249}.

In summary, this example shows why \df compares pre-commit and post-commit slices. Fuzzing the post-commit slice can expose the overflow, but the pre-commit slice provides the baseline needed to determine whether the bug was introduced by the commit. In this example, the pre-commit guard blocks \texttt{descsz > 20}, while the post-commit guard allows such values to reach \texttt{memcpy}. 
By comparing the paired sanitizer reports, \df identifies the overflow as a candidate commit-induced bug without whole-program fuzzing or a manually written test case.

\section{\df Approach}
\label{sec}

\begin{figure}[tb!]
\centering
\includegraphics[width=0.95\columnwidth]{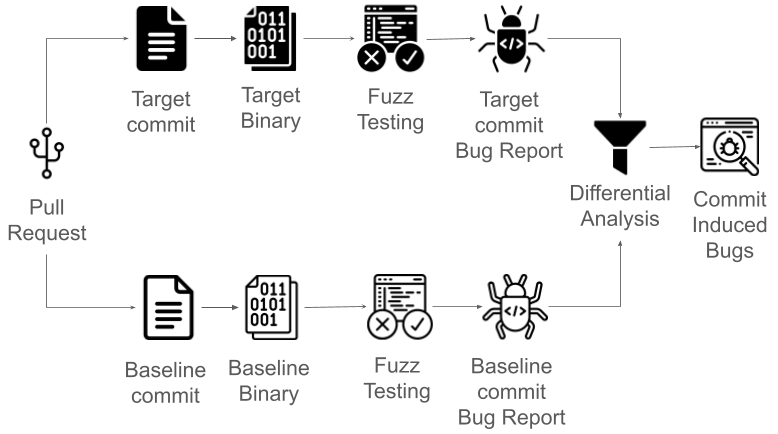}
\caption{Overview of \df.}

% Given a target commit and a baseline commit, \df identifies modified functions, constructs paired pre-commit and post-commit slices, fuzzes both slices, and reports bugs that appear only after the commit.}
\label{img}
\end{figure}

\df verifies code changes by formulating commit-level analysis as a differential slice-fuzzing problem. Given a target commit and a baseline commit from the same repository, such as the immediate predecessor of the target commit, \df checks whether the target commit introduces new memory-safety bugs. The key idea is to compare the behavior of paired executable slices, i.e., one extracted from the baseline version and one extracted from the target version. A sanitizer report is treated as a candidate commit-induced bug only if it appears in the post-commit slice and not in the corresponding pre-commit slice. 
Figure~\ref{img} shows an overview of \df.

\subsection{Inputs and Definitions}
\df is  triggered on individual commits or pull requests, enabling \textit{on-demand} slice fuzzing for commit verification.
In the case of commits, both a target commit and a baseline commit are required. 
The target commit is the version of code under verification. 
The baseline commit provides the reference version used for comparison. In our evaluation, the baseline is the immediate predecessor of the target commit. 
The same formulation can be used for pull requests by treating the pre-merge and post-merge versions as the paired commits. These paired commits are the primary input for the approach.

% For each modified function in the code changes, \df constructs two executable artifacts. The pre-commit slice is extracted from the baseline version, while the post-commit slice is extracted from the target version.
% These paired slices contain the modified function and the dependencies needed to compile and fuzz it in isolation. Fuzzing each slice produces sanitizer reports, each associated with a bug class and a source location.
% \df compares these reports across the paired slices to identify reports that are newly introduced by the target commit.

\subsection{Identifying Modified Functions}

\df first identifies the relevant functions between the paired commits. 
It computes the diff between the target commit and the baseline commit using the project's version-control history. 
From this diff, \df extracts the source files and functions that were modified between the two versions.

\df focuses on modified functions because they provide code in both versions, which enables differential comparison. Deleted functions cannot execute in the target commit and are therefore excluded.
Newly added functions do not have corresponding implementations in the baseline version, so they cannot be analyzed using the same differential oracle. 
We therefore exclude added functions from the differential analysis and discuss this limitation in Section~\ref{sec:limitation}.

To compare reports across versions, \df also records a mapping between source locations in the target commit and corresponding locations in the baseline commit. This mapping is derived from the hunk information in the diff.
It allows \df to determine whether a sanitizer report in the post-commit slice corresponds to a report at the same mapped location in the pre-commit slice.

\subsection{Constructing Paired Code Slices}
For each modified function, \df constructs paired code slices from the baseline and target versions.
Each slice contains the function under analysis and the dependencies required to compile it as an executable fuzzing target. 
This slice-level design avoids fuzzing the entire program while still preserving enough context to execute the modified function and its relevant dependencies.

The important requirement is comparability. The pre-commit and post-commit slices must represent the same function across versions so that \df can interpret differences in sanitizer reports as changes caused by the commit. When a modified function exists in both versions, \df extracts and compiles the corresponding slices independently. These paired slices form the basis for the differential analysis.

\subsection{Fuzzing Paired Slices}

After constructing the paired slices, \df generates a fuzzing wrapper for each slice and links the wrapper with the compiled slice objects to produce executable fuzzing binaries. For every modified function, this results in two binaries: one for the baseline slice and one for the target slice.

\df fuzzes both binaries independently. During fuzzing, runtime sanitizers act as bug oracles and report memory-safety issues such as buffer overflows, memory leaks, uses of uninitialized memory, and undefined behavior. Each report records the bug class and the source location where the sanitizer detects the failure. These reports are not immediately shown to developers. Instead, they are passed to the differential oracle described next.

\subsection{Differential Bug Oracle}

The differential oracle is the core of \df. 
Fuzzing an isolated slice can expose real bugs, but it can also expose behavior that is not specific to the commit, including pre-existing bugs or artifacts caused by relaxed slice execution. \df therefore interprets a sanitizer report relative to the corresponding pre-commit slice.

A post-commit sanitizer report is classified as a candidate commit-induced bug only when all of the following conditions hold. 
First, the report appears in the post-commit slice. 
Second, no report with the same bug class appears at the corresponding mapped location in the pre-commit slice. Third, the mapped location is reachable in the pre-commit slice, which helps distinguish newly introduced behavior from differences caused only by missing coverage in the baseline version.

This oracle allows \df to filter reports that are already present before the commit and prioritize reports that newly appear after the code change. The output of \df is therefore not the full set of sanitizer reports produced by slice fuzzing. Instead, \df reports a smaller set of candidate commit-induced bugs that are more relevant to code review and manual triage.

\section{Implementation}
\label{sec}

This section describes how \df implements the differential slice-based fuzzing method.
The implementation has four main components: commit-aware slice extraction, fuzzing-wrapper generation, binary instrumentation, and parallel fuzzing with input replay.
Our prototype uses \oldfs~\cite{FuzzSlice} as the slicing and compilation backend, while \df adds the commit-level pairing, differential execution, and reporting logic needed for commit verification.

\subsection{Commit-Aware Slice Extraction}

\df uses the modified functions identified from the target and baseline commits as slicing targets. For each modified function that exists in both versions, \df utilizes \oldfs~\cite{FuzzSlice} to extract a source-level slice and compiles it into object files and binaries. Each slice includes the target function and the dependencies required to compile and execute it in isolation. \df performs this process separately for the baseline and target versions, producing paired slice binaries for each modified function.

% Differential comparison requires comparable code in both versions. \df therefore excludes functions that are newly added or deleted by the target commit. Deleted functions cannot execute in the target version, while newly added functions do not have a corresponding baseline implementation. We discuss this with more detail in  Section~\ref{sec:limitation}.

\subsection{Fuzzing Wrapper Generation}

Each compiled slice requires a fuzzing wrapper that initializes the function inputs and invokes the target function. \df generates wrappers from function parameter types. The wrapper allocates memory for pointer fields, initializes nested structures, populates scalar values, and exposes these inputs to the fuzzer.

Listing~\ref{lst-wrapfuzz} shows an example wrapper for a function \texttt{process\_buffer} that operates on a \texttt{buffer} structure. The wrapper allocates a \texttt{buffer} object, allocates memory for its pointer field \texttt{data}, and initializes the \texttt{len} field. These fields are populated from fuzz input using \texttt{FUZZ\_DATA}. After randomly initializing the size of \texttt{data}, the wrapper copies fuzz-generated bytes into \texttt{data} and invokes the target function \texttt{process\_buffer}. This setup enables the fuzzer to explore program behaviors that depend on the input buffer contents.

\df optimizes wrapper generation by initializing only fields that are used by the slice. To identify these fields, \df analyzes the sliced code using \texttt{tree-sitter}~\cite{treesitter}. If a structure contains fields that are never accessed by the target slice, \df omits their initialization. This reduces the number of fuzz bytes needed to construct inputs and focuses mutation effort on values that can affect slice execution.

\begin{lstlisting}[language=C,
caption={Example fuzzing wrapper with nested structures.},
label={lst-wrapfuzz},
breaklines=true,
numbers=left,
numberstyle=\tiny,
xleftmargin=2em,
float=tb,
basicstyle=\ttfamily\footnotesize,
commentstyle=\ttfamily\footnotesize,
escapeinside={(@}{@)}
]
void process_buffer(struct buffer *b);
struct buffer {
    char *data;
    int len;
};

int main(char *Fuzz_Data, size_t Fuzz_Size) {
    struct buffer *b = malloc(sizeof(struct buffer));
    int alloc_size;
    FUZZ_DATA(&alloc_size);
    b->data = malloc(alloc_size);
    memset(b->data, 0, alloc_size);
    FUZZ_DATA(b->data);
    FUZZ_DATA(&b->len);
    process_buffer(b);
}
\end{lstlisting}

\subsection{Binary Construction and Instrumentation}

For each slice, \df links the generated wrapper with the compiled slice object files to produce an executable fuzzing binary. For every modified function, this produces two binaries: one from the baseline slice and one from the target slice. 
These paired binaries are then fuzzed and compared by the differential oracle.

\df instruments each binary with runtime sanitizers to detect memory-safety bugs. Because some sanitizers are incompatible when enabled together, \df builds two instrumented variants for each slice. The first combines AddressSanitizer, LeakSanitizer, and UndefinedBehaviorSanitizer~\cite{serebryany2012addresssanitizer, leaksanitizer, UBSAN}. The second uses MemorySanitizer~\cite{MSAN} and coverage instrumentation. Inputs generated during fuzzing are replayed on both variants, enabling \df to detect buffer overflows, memory leaks, uses of uninitialized memory, integer overflows, and other undefined behaviors while also collecting coverage information.

\subsection{Fuzzing and Input Replay}

\df fuzzes each generated binary using AFL++~\cite{afl++}. Since fuzzing is randomized, the baseline and target slices may explore different paths even when they are fuzzed with the same time budget. This can affect differential comparison because a report may appear absent from one version simply because the relevant path was not exercised.

To reduce this source of noise, \df shares generated test cases across paired slices. Inputs discovered while fuzzing the target slice are replayed on the baseline slice, and inputs discovered while fuzzing the baseline slice are replayed on the target slice. This replay step makes the comparison more reliable by increasing the chance that both versions are evaluated on the same input behaviors. It also helps distinguish reports that are truly introduced by the target commit from reports that arise due to uneven fuzzing exploration.

\subsection{Parallel Execution and Reporting}

\df runs slicing, compilation, fuzzing, and replay tasks in parallel using a Kubernetes-based execution pipeline. Each modified function and each commit version can be processed independently, which allows \df to analyze multiple commits and functions concurrently. This design supports the commit-level use case, where developers need feedback quickly enough to inform code review.

For each analyzed slice, \df records sanitizer reports, coverage information, and representative stack traces. After differential filtering, \df reports only candidate commit-induced bugs to developers. For each reported bug, \df provides the bug class, the source location, the corresponding commit, and up to three representative stack traces to support manual triage.

\section{Evaluation}
\label{sec:results}

We evaluate \df through the following RQs:

\begin{itemize}
[leftmargin=1.3em, itemsep=0pt, parsep=0pt, topsep=2pt]
\item[-] \textbf{RQ1:} \RQone
\item[-]  \textbf{RQ2:} \RQtwo
\item[-] \textbf{RQ3:} \RQthree
\item[-] \textbf{RQ4:} \RQfour
\end{itemize}

The evaluation first describes the benchmark projects, the commit selection process, and the experimental configuration. It then reports \df's effectiveness in reducing raw sanitizer reports to candidate commit-induced bugs, its runtime and coverage, the confirmed bugs found across projects, and the causes of the remaining false positives.

\subsection{Evaluation Setup}
\label{setup}

\begin{table}[tb]
  \caption{Statistics on the repositories in our benchmark.}
  \centering
  \begin{tabular}{lrrr}
    \toprule
    \textbf{Repository} & \textbf{Lines of Code} & \textbf{Commits Analyzed} & \textbf{Start Date} \\
    \midrule
    libpcap    & 39k & 100 & Jul 24, 2025 \\
    OpenSSL    & 620k & 100 & Oct 15, 2025 \\
    leptonica  & 191k & 100 & Mar 19, 2025 \\
    \bottomrule
  \end{tabular}
  \label{Repo stats}
\end{table}

\noindent
\textbf{Dataset.}
\df is designed to verify individual commits by analyzing functions modified by a code change. To evaluate \df, we analyze commits from three widely used real-world projects: \href{https://github.com/openssl/openssl}{OpenSSL}, \href{https://github.com/the-tcpdump-group/libpcap}{libpcap}, and \href{https://github.com/DanBloomberg/leptonica}{leptonica}. Table~\ref{Repo stats} summarizes the selected projects which cover different application domains. OpenSSL is a widely deployed cryptographic library that provides secure communication primitives such as TLS/SSL. libpcap is a low-level packet-capture library used for network traffic inspection. leptonica is an image-processing library used in computer vision. 
% Together, these projects provide a diverse benchmark for evaluating whether \df can support commit-level verification across different kinds of systems code.

For each project, we analyze 100 commits from the release branch, starting from the date shown in Table~\ref{Repo stats}. Each commit is treated as a target commit, and its immediately preceding commit is used as the baseline. For each target-baseline pair, \df identifies modified functions, constructs paired pre-commit and post-commit slices, fuzzes both slices independently, and compares the resulting sanitizer reports using the differential oracle. Reports that remain after differential filtering are manually inspected. 
Confirmed reports are submitted to the corresponding project maintainers, while rejected reports are counted as false positives.
For double-blind review, we omit links to public bug reports and developer discussions because they reveal author identities. These links will be restored upon the publication of the paper.

\noindent
\textbf{Configuration.}
We allocate a fuzzing budget of 3 minutes per slice. Experiments are performed on Ubuntu. Some functions are skipped when their slices cannot be built or executed on this platform. The experiments run on a bare-metal Kubernetes cluster with two AMD EPYC 9224 worker nodes, each with 48 cores at 2.5 GHz and 1 TB of RAM. Each fuzzing job is pinned to one core and allocated 20 GB of memory, with up to 20 jobs running concurrently.

\begin{table}[tb!]
\centering
\caption{Summary of Differential Analysis Across Projects}
\label{tab:analysis-summary}
\resizebox{\columnwidth}{!}{%
\begin{tabular}{lcccccc}
\toprule
\textbf{Project} & \makecell{\textbf{Commits}} & \makecell{\textbf{Functions}} & \makecell{\textbf{Commits} \\ \textbf{with Diffs}} & \makecell{\textbf{Bugs}} & \makecell{\textbf{True} \\ \textbf{Bugs}} & \makecell{\textbf{False} \\ \textbf{Bugs}} \\
\midrule
OpenSSL   & 100  & 122 & 1 & 1 & 1 & 0 \\
libpcap   & 100  & 141 & 4 & 4 & 2  & 2 \\
leptonica & 100 & 144 & 2 & 2 & 2 & 0 \\
\midrule
Total     & 300 & 407 & 7 & 7 & 5 & 2 \\
\bottomrule
\end{tabular}%
}
\end{table}

\subsection{RQ1: \RQone}
\label{res1}

Table~\ref{tab:analysis-summary} summarizes the results of \df's differential analysis. Across 300 commits, \df analyzes 407 modified functions. Fuzzing these slices initially produces 518 sanitizer reports before differential filtering. These raw reports include all failures exposed by isolated slice fuzzing and therefore include pre-existing bugs and slice level failures that are not necessarily introduced by the target commit.

\df then applies its differential oracle by comparing sanitizer reports from paired pre-commit and post-commit slices. A report is retained only when it appears in the post-commit slice and not in the corresponding pre-commit slice. This filtering step reduces the 518 raw reports to 7 candidate commit-induced bug reports across 7 commits. Manual validation confirms 5 of these reports as real bugs and classifies 2 as false positives.

The confirmed bugs are distributed across all three projects.
OpenSSL contains 1 confirmed bug, leptonica contains 2 confirmed bugs, and libpcap contains 2 confirmed bugs and 2 false positives. These results show that \df can substantially reduce the triage burden for developers while preserving reports that correspond to real commit-induced bugs.

\begin{figure}[tb!]
\centering
\resizebox{\columnwidth}{!}{
\begin{tikzpicture}
\pie[
    text=legend,
    hide number,
    radius=3,
    sum=auto,
    color={
        blue!60!white, red!60!white, green!60!white, orange!60!white, purple!60!white,cyan!60!white, magenta!60!white, yellow!70!white,teal!60!white, lime!60!white, brown!60!white, pink!60!white,gray!50!white, violet!60!white, olive!60!white, black!30!white,red!50!white, blue!50!white,
    }
]
{
    25.77/Segmentation fault (25.77\%),
    20.03/Buffer overflow (20.03\%),
    16.68/Null pointer deref (16.68\%),
    12.72/Misaligned memory access (12.72\%),
    9.06/Integer overflow (9.06\%),
    6.31/Use-after-free (6.31\%),
    1.1/Uninitialized memory (1.10\%),
    1.87/Unrepresentable computed value (1.87\%),
    0.09/Testcase timeout (0.09\%),
    0.69/Malloc too much memory (0.69\%),
    3.35/Memory Leak (3.35\%),
    0.49/Incorrect function call (0.49\%),
    0.29/Stack overflow (0.29\%),
    0.49/Double free (0.49\%),
    0.09/Abrupt exit (0.09\%),
    0.19/Division by zero (0.19\%),
    0.29/Floating point exception (0.29\%),
    0.49/Bit Shift Undefined Behavior (0.49\%)
}
\end{tikzpicture}
}
\caption{Distribution of bug types across slices before differential analysis.}
\label{fig:bug-types}
\end{figure}
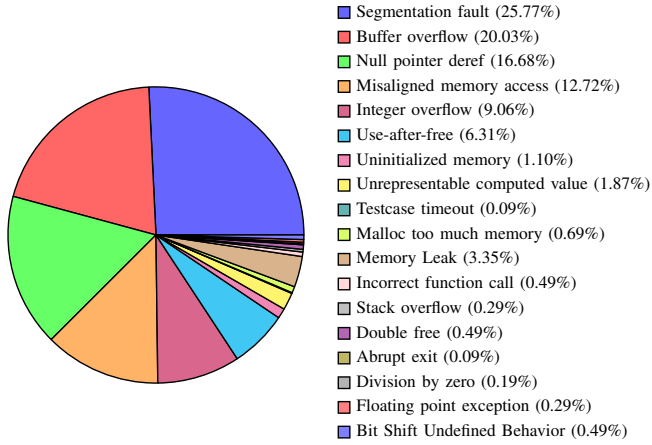

Figure~\ref{fig:bug-types} shows the distribution of sanitizer reports before differential filtering. The raw reports include several memory-safety failure classes, including segmentation faults, buffer overflows, NULL pointer dereferences, misaligned memory accesses, integer overflows, use-after-free reports, and memory leaks. This distribution shows that slice fuzzing can expose a broad range of failures, but it also highlights why differential filtering is necessary. Without the pre-commit baseline, developers would need to inspect hundreds of raw reports that are not necessarily caused by the target commit.

\begin{tcolorbox}[colback=gray!5,colframe=gray!60,boxrule=0.4pt,left=1mm,right=1mm,top=1mm,bottom=1mm]
\noindent
\textbf{Answer.} 
\df reduces 518 raw sanitizer reports to 7 candidate commit-induced bug reports, of which 5 are confirmed as true bugs and 2 are classified as false positives.

\noindent\textbf{Insight.}
Differential filtering is central to our approach, as isolated slice fuzzing exposes many failures, but comparing pre-commit and post-commit slices turns this noisy output into a small set of reports relevant to the target commit.
\end{tcolorbox}

%%%%%===========++RQ2====================

\begin{table}[tb!]
\centering
\caption{Analysis Time Across Projects}
\label{tab:analysis-time}
\resizebox{\columnwidth}{!}{%
\begin{tabular}{lccccc}
\toprule
\textbf{Project} & \textbf{\# Commits} & \textbf{\# Functions} & \textbf{Total Time} & \makecell{\textbf{Avg. Time} \\ \textbf{per Commit}} & \makecell{\textbf{Avg. Time} \\ \textbf{per Function}} \\
\midrule
OpenSSL & 100  & 122  & 58.3h & 35m & 28.6m \\
libpcap & 100  & 141  & 46.6h & 28m & 20m \\
leptonica & 100 & 144 & 57.6h & 34.2m & 24.5m \\
\midrule
Average & - & - & 54.1h & 32.4m & 24.1m \\
\bottomrule
\end{tabular}%
}
\end{table}

\subsection{RQ2: \RQtwo}
\label{res2}

Table~\ref{tab:analysis-time} reports the time required to analyze each project. Across all projects, \df requires 32.4 minutes per commit and 24.1 minutes per modified function on average. This runtime includes slice construction, binary generation, fuzzing, input replay, sanitizer execution, and report collection.

The total analysis time varies across projects. OpenSSL requires 58.3 hours to analyze 100 commits and 122 modified functions, corresponding to 35 minutes per commit and 28.6 minutes per function. leptonica requires 57.6 hours to analyze 100 commits and 144 modified functions, corresponding to 34.2 minutes per commit and 24.5 minutes per function. libpcap is faster, requiring 46.6 hours to analyze 100 commits and 141 modified functions, corresponding to 28 minutes per commit and 20 minutes per function.

The higher analysis times for OpenSSL and leptonica are mainly caused by longer build and slice-construction times. These costs are paid for each analyzed version because \df needs build information to construct executable slices. Differential or incremental build reuse could reduce this overhead in future implementations. Even with this cost, the observed turnaround time is practical for commit-level analysis because the tool produces a small set of candidate reports within a time scale that can support code review.

\begin{table}[tb!]
\centering
\caption{Coverage Metrics Across Projects}
\label{tab:coverage-summary}
\resizebox{\columnwidth}{!}{%
\begin{tabular}{lccccc}
\toprule
\textbf{Project} & \textbf{\# Commits} & \textbf{\# Functions} & \textbf{Function Cov.} & \textbf{Slice Cov.} & \textbf{Total Cov.} \\
\midrule
% Example rows (replace with your data)
OpenSSL & 100  & 122  & 73.94\% & 52.71\% & 71.68\% \\
libpcap & 100  & 141  & 79.89\% & 52.85\% & 80.18\% \\
leptonica & 100 & 144 & 72.35\% & 19.75\% & 24.43\% \\
\midrule
Average & - & - & 75.36\% & 41.69\% & 58.62\% \\
\bottomrule
\end{tabular}%
}
\end{table}

Table~\ref{tab:coverage-summary} reports coverage achieved within the 3-minute fuzzing budget per slice. We report three coverage metrics. Target function coverage measures how much of the modified function is executed. Slice coverage measures how much of the extracted slice is executed, including dependencies. Total coverage measures coverage over the slice and generated wrapper together.

Across all projects, \df covers 75.36\% of modified functions and 41.69\% of extracted slices on average. The modified-function coverage is important because commit-level verification depends on exercising the changed code. OpenSSL achieves 73.94\% function coverage, 52.71\% slice coverage, and 71.68\% total coverage. 
libpcap achieves the highest modified-function coverage, with 79.89\% function coverage, 52.85\% slice coverage, and 80.18\% total coverage.
leptonica achieves 72.35\% function coverage, but its slice and total coverage are lower, at 19.75\% and 24.43\%, respectively. 

The lower slice coverage in leptonica is due to excessive function pointers, which lead to larger slices with numerous reachable functions. 
These larger slices are harder to cover within a short fuzzing budget. 
Overall, these coverage results show that \df exercises a substantial portion of modified functions while maintaining a short analysis budget.

\begin{tcolorbox}[colback=gray!5,colframe=gray!60,boxrule=0.4pt,left=1mm,right=1mm,top=1mm,bottom=1mm]
\noindent\textbf{Answer.}
\df analyzes a commit in 32.4 minutes on average and covers 75.36\% of modified functions within a 3-minute fuzzing budget per slice.

\noindent\textbf{Insight.}
The changed function is usually exercised quickly by \df, making differential slice fuzzing suitable for commit-level review rather than only for long-running offline testing.
\end{tcolorbox}

\subsection{RQ3: \RQthree}
\label{res3}

We now describe the confirmed bugs identified by \df through differential analysis. 
These cases illustrate how \df detects bugs introduced or exposed by code changes across different projects and bug types.
Below, we describe in detail the bug types confirmed in each project.

\begin{lstlisting}[language=C, 
  caption={Example of a commit inducing a buffer overflow within OpenSSL~\cite{openssl-buffer-overflow-commit}.}, 
  label={lst:openssl-buffer-overflow},
  breaklines=true,
  numbers=left,
  numberstyle=\tiny,
  xleftmargin=2em,
  basicstyle=\ttfamily\footnotesize,
  keywordstyle=\bfseries,
  escapeinside={(*@}{@*)}
]
static void put_str(const char *str, char **buf, size_t *remain, size_t *needed)
{
  size_t olen, len, i; int quotes;
  char quote = '\0';
  len = olen = strlen(str);
  *needed += len;
(*@\fulldiffdel{- if (*remain == 0) return;}@*)
  for (i = 0; i < len; i++)
      if (str[i] != '.' && str[i] != '_') {
          if (quote == '\0') quote = '\'';
          if (str[i] == '\'') quote = '"';
  }
  quotes = quote != '\0';
(*@\fulldiffadd{+ if (*remain == 0) \{}@*)
(*@\fulldiffadd{+ *needed += 2 * quotes;}@*)
(*@\fulldiffadd{+ return;}@*)
(*@\fulldiffadd{+ \}}@*)
(*@\fulldiffadd{+ if (quotes)}@*)
(*@\fulldiffadd{+   put\_char(quote, buf, remain, needed);}@*)
(*@\fulldiffdel{- if (*remain < len + 1)}@*)
(*@\fulldiffadd{+ if (*remain < len + 1 + quotes)}@*)
      len = *remain - 1;
  if (len > 0) memcpy(*buf, str, len);
}
\end{lstlisting}

\noindent
\textbf{Buffer overflow in OpenSSL.}
Listing~\ref{lst:openssl-buffer-overflow} shows a commit that introduces a buffer overflow in OpenSSL. The commit modifies \texttt{put\_str}, which copies quoted strings into \texttt{buffer}. The function tracks the number of remaining bytes in the destination buffer using the unsigned integer variable \texttt{remain}.

Before the change, the function returned early when \texttt{remain == 0}, preventing any copy when no space was available. The commit moves this check after quote detection so that the function can account for opening and closing quotes. However, this modified logic creates an edge case when the input requires quoting and only one byte remains in the destination buffer.

In this case, the opening quote is copied first, reducing \texttt{remain} to 0. The subsequent computation \texttt{remain - 1} then underflows because \texttt{remain} is unsigned. This produces a large value, allowing too much data to be copied into \texttt{buffer} and causing a buffer overflow. \df detects the overflow in the post-commit slice but not in the corresponding pre-commit slice. 
Upon reporting this to the project developers, they fixed the issue, credited the authors, and added targeted tests for this edge condition.
As mentioned earlier, we omit links to public bug reports and developer discussions because they reveal author identities.

%~\cite{openssl-buffer-overflow-fix}.

\noindent
\textbf{Memory leak in leptonica.}
Listing~\ref{lst:leptonica-mem-leak} shows a commit that introduces a memory leak in leptonica. The commit modifies \texttt{jbDataRender}, which reconstructs full page images from compressed data. The original code creates a \texttt{box}, renders it onto an image, and then explicitly deallocates it.

The new code changes ownership behavior by inserting the created \texttt{box} into a list before rendering all boxes. The list is later destroyed and should release its contained boxes. However, \df generates an input that causes insertion into the list to fail. When this failure occurs, ownership of the newly created \texttt{box} is not transferred to the list, and the object is not explicitly deallocated. This produces a memory leak in the post-commit slice that does not appear in the paired pre-commit slice. The project developers responded to our report on this bug and clarified the ownership contract and later fixed leaks in similar locations within the project.
% ~\cite{leptonica-mem-leak-fix}.

\begin{lstlisting}[language=C, 
  caption={Example of a commit inducing a memory leak within leptonica~\cite{leptonica-mem-leak-commit}.}, 
  label={lst:leptonica-mem-leak},
  breaklines=true,
  numbers=left,
  numberstyle=\tiny,
  xleftmargin=2em,
  float=tb,
  basicstyle=\ttfamily\footnotesize,
  keywordstyle=\bfseries,
  escapeinside={(*@}{@*)}
]
PIXA *jbDataRender(JBDATA  *data, l_int32  debugflag) {
  ...
  box = boxCreate(x, y, wp, hp);
(*@\fulldiffdel{-   /* Use box */}@*)
(*@\fulldiffdel{-   pixRenderBoxArb(pixd, box, 1, 255, 0, 0);}@*)
(*@\fulldiffdel{-   /* Deallocate box */}@*)
(*@\fulldiffdel{-   boxDestroy(\&box);}@*)
(*@\fulldiffadd{+   /* Add box to list of boxes */}@*)
(*@\fulldiffadd{+   boxaAddBox(boxa, box, L\_INSERT);}@*)
(*@\fulldiffadd{+   /* Destroy list of boxes */}@*)
(*@\fulldiffadd{+   boxaDestroy(\&boxa);}@*)
}
\end{lstlisting}

\noindent
\textbf{Buffer overread and NULL pointer dereference in libpcap.}
Listing~\ref{lst} shows a libpcap commit that exposes a buffer overread and a subsequent NULL pointer dereference. The function \texttt{newchunk\_nolongjmp} allocates memory chunks from a shared context, \texttt{cstate}. The key defect is that the function accesses the current chunk before validating that the updated chunk index remains within the valid chunk array.

The relevant commit modifies \texttt{gen\_port}. Before the change, the function returned the constructed block directly. After the change, it passes the block to \texttt{gen\_port\_common}, introducing an additional execution path that performs more allocations through \texttt{newchunk\_nolongjmp}. Under an input generated by \df, the post-commit slice exhausts the available chunk list and reaches the invalid access before the bounds check. Once exhaustion is detected, the allocator returns \texttt{NULL}; downstream code then fails to handle this value correctly, producing a NULL pointer dereference. This case illustrates that a commit can expose a latent defect by changing reachability or allocation behavior. 
This led to multiple libpcap fixes, with the authors being credited for the finding.
% The finding led to multiple fixes in libpcap, and the authors were credited for identifying the issue.
% ~\cite{libpcap-buffer-overread-fix, libpcap-buffer-overread-pr}.

\begin{lstlisting}[language=C,
caption={Example of a commit exposing a buffer overread and NULL dereference in libpcap~\cite{libpcap-buffer-overread-commit}.},
label={lst},
breaklines=true,
numbers=left,
numberstyle=\tiny,
xleftmargin=2em,
float=tb,
basicstyle=\ttfamily\footnotesize,
keywordstyle=\bfseries,
escapeinside={(@}{@)}
]
void *newchunk_nolongjmp(compiler_state_t *cstate, size_t n) {
  struct chunk *cp; int k;
  n = (n + CHUNK_ALIGN - 1) & ~(CHUNK_ALIGN - 1);
  /* Access memory before checking bounds */
  cp = &cstate->chunks[cstate->cur_chunk];
  if (n > cp->n_left) {
    ++cp;
    k = ++cstate->cur_chunk;
    /* Checking bounds */
    if (k >= NCHUNKS) {
      bpf_set_error(cstate, "out of memory");
      return NULL;
    }
  }
  /* allocate and initialize the new chunk */
  ...
}

struct block *gen_port(compiler_state_t *cstate,
uint16_t port, int proto, int dir) {
  struct block *b1, *tmp;
  tmp = gen_portatom(cstate, 0, port);
  b1 = gen_portatom(cstate, 2, port);
  gen_or(tmp, b1);
(@\fulldiffdel{- return b1;}@)
(@\fulldiffadd{+ return gen\_port\_common(cstate, proto, b1);}@)
}
\end{lstlisting}

\noindent
\textbf{Integer overflow in leptonica.}
Listing~\ref{lst:leptonica-integer-overflow} shows a leptonica commit that exposes an integer overflow. The function \texttt{numaGetPeakCentroids} computes centroids between successive histogram peaks. It uses \texttt{nahist} to store histogram values and \texttt{narange} to store index ranges for detected peaks. The function iterates over \texttt{narange}, extracts a pair of indices, \texttt{low} and \texttt{high}, and then traverses the corresponding histogram interval to compute a weighted centroid.
The commit prevents Division-By-Zero (DBZ) by adding a guard when computing the centroid.
However, another bug persists unfixed by the commit: \df can create an invalid range in which the loop bounds do not correspond to a valid histogram interval.
This can cause the loop variable \texttt{j} to overflow while iterating from \texttt{low} to \texttt{high}, leading to undefined behavior.
By eliminating DBZ failure condition, the new guard enables a larger set of inputs to progress further through execution, increasing the opportunity for loop variable \texttt{j} to overflow. 
Differential fuzzing detects this behavioral difference through overflow reports that appear exclusively in the post-commit version.
The issue was reported and fixed by the developers.
% ~\cite{leptonica-integer-overflow-fix}.

\begin{tcolorbox}[colback=gray!5,colframe=gray!60,boxrule=0.4pt,left=1mm,right=1mm,top=1mm,bottom=1mm]
\textbf{Answer.} \df detects several classes of commit-induced memory-safety bugs, including buffer overflows, memory leaks, buffer overreads, NULL pointer dereferences, and integer overflows.

\noindent\textbf{Insight.}
The detected bugs are not limited to direct unsafe edits; some commits expose latent defects by changing control flow, allocation behavior, or reachability.
\end{tcolorbox}

\begin{lstlisting}[language=C, 
  caption={Example of a commit exposing an integer overflow in leptonica~\cite{leptonica-integer-overflow-commit}.},
  label={lst:leptonica-integer-overflow},
  breaklines=true,
  numbers=left,
  numberstyle=\tiny,
  xleftmargin=2em,
  basicstyle=\ttfamily\footnotesize,
  keywordstyle=\bfseries,
  escapeinside={(*@}{@*)}
]
static NUMA* numaGetPeakCentroids(NUMA *nahist, NUMA *narange) {
  l_int32    i, j, nr, low, high;
  l_float32  cent, sum, val; NUMA *nad;
  nr = numaGetCount(narange) / 2;
  nad = numaCreate(4);
  for (i = 0; i < nr; i++) {
      numaGetIValue(narange, 2 * i, &low);
      numaGetIValue(narange, 2 * i + 1, &high);
      cent = sum = 0.0;
      for (j = low; j <= high; j++) {
          numaGetFValue(nahist, j, &val);
          cent += j * val;
          sum += val;
      }
(*@\fulldiffadd{+ if (sum > 0) numaAddNumber(nad, cent / sum);}@*)
(*@\fulldiffdel{- numaAddNumber(nad, cent / sum);}@*)
  }
  return nad;
}
\end{lstlisting}

\subsection{RQ4: \RQfour}
\label{res4}

Although our differential approach substantially reduces the number of reports, \df still produces 2 false positives. We analyze these cases to understand the limitations of differential slice-based fuzzing.

\noindent
\textbf{False positive caused by a missing full-program precondition.} Listing~\ref{lst-fp1} shows the first false positive, which occurs in libpcap in \texttt{gen\_portrange6}. This function constructs a condition that checks whether an IPv6 packet's port number falls within a specified range and whether the protocol is allowed. Before the change, the function enforced the protocol check through \texttt{gen\_ip6\_proto} and then combined it with the port-range check produced by \texttt{gen\_portrangeatom6}.

\begin{lstlisting}[language=C,
caption={False positive NULL dereference caused by a missing full-program precondition in libpcap~\cite{libpcap-buffer-overread-commit}.},
label={lst-fp1},
breaklines=true,
numbers=left,
numberstyle=\tiny,
xleftmargin=2em,
float=tb,
basicstyle=\ttfamily\footnotesize,
keywordstyle=\bfseries,
escapeinside={(@}{@)}
]
struct block *gen_portrange6(compiler_state_t *cstate, int port1, int port2, int proto) {
struct block *b0, *b1, *tmp;
(@\fulldiffdel{- b0 = gen\_ip6\_proto(cstate, proto);}@)
  b1 = gen_portrangeatom6(cstate, port1, port2);
(@\fulldiffdel{- gen\_and(b0, b1);}@)
return b1;
}
\end{lstlisting}

The commit removes the local protocol-filtering logic, reflecting a design change in which protocol filtering is handled elsewhere. In the isolated post-commit slice, \df constructs an input that reaches \texttt{gen\_portrangeatom6} with a NULL \texttt{cstate}, producing a NULL pointer dereference. 
Our differential analysis highlights this as a new bug because a NULL cstate in the pre-commit slice can only trigger a NULL dereference at \texttt{gen\_ip6\_proto}.
However, in the full program, \texttt{gen\_portrange6} is not intended to be invoked with a NULL \texttt{cstate}. The report, therefore, violates an implicit precondition enforced by the broader program context and does not represent a realizable bug. This false positive shows that isolated slices can over-approximate valid inputs when important caller-side invariants are missing.

\noindent
\textbf{False positive caused by wrapper mismatch.}
Listing~\ref{lst-fp2} shows the second false positive, which occurs after a function-signature change in libpcap. The commit modifies \texttt{gen\_proto} by removing the argument \texttt{dir} and deleting the corresponding check. This change does not alter the intended semantics of the function and should not introduce a memory-safety bug.

\begin{lstlisting}[language=C,
caption={False positive caused by wrapper mismatch after a function-signature change in libpcap~\cite{libpcap-fp-oob-commit}.},
label={lst-fp2},
breaklines=true,
numbers=left,
numberstyle=\tiny,
xleftmargin=2em,
float=tb,
basicstyle=\ttfamily\footnotesize,
keywordstyle=\bfseries,
escapeinside={(@}{@)}
]
static int alloc_reg(compiler_state_t *cstate) {
int n = cstate->curreg++;
...
}
struct block *gen_proto(compiler_state_t *cstate,
(@\fulldiffdel{- int dir,}@)
int protochain) {
(@\fulldiffdel{- if (dir != Q\_DEFAULT)}@)
(@\fulldiffdel{- return gen\_true(cstate);}@)
...
}
\end{lstlisting}

\df reports an out-of-bounds read in the post-commit slice in the transitively called function \texttt{alloc\_reg}. The same issue is not triggered in the pre-commit slice, even though the relevant code region is covered. Manual inspection shows that the false positive is caused by inconsistent exploration between the pre-commit and post-commit slices. \df normally mitigates this problem by replaying inputs across paired slices, but in this case the function-signature change causes the automatically generated wrappers to differ. As a result, the same input bytes are interpreted differently across versions, reducing the effectiveness of input replay.

\begin{tcolorbox}[colback=gray!5,colframe=gray!60,boxrule=0.4pt,left=1mm,right=1mm,top=1mm,bottom=1mm]
\noindent
\textbf{Answer.} 
\df produces 2 false positives: one caused by a missing full-program precondition and one caused by wrapper mismatch after a function-signature change.

\noindent\textbf{Insight.}
The false positives come from limits of isolated slice execution, especially when caller-side invariants or stable input encodings are missing.
\end{tcolorbox}

\section{Limitations and Future Work}
\label{sec:limitation}
This section presents the limitations and potential future research opportunities for improving \df.

\noindent
\textbf{Newly added functions are not directly analyzed.}
\df relies on comparing paired pre-commit and post-commit slices. Therefore, it directly analyzes modified functions that exist in both versions. Newly added functions do not have a corresponding pre-commit implementation, so \df cannot apply the same differential oracle to them. As a result, bugs confined entirely to newly added functions may be missed. However, if newly added code affects existing modified functions, \df can still detect failures that manifest through those modified functions. In future work, we plan to extend \df to analyze callers of newly added functions, allowing the framework to reason about new code through existing call paths.

\noindent
\textbf{Isolated slices may miss full-program invariants.}
Because \df fuzzes executable slices rather than the full program, some full-program preconditions may be absent from the slice. This can cause false positives when the fuzzer generates inputs that are valid for the isolated slice but not realizable in the full program. Our RQ4 analysis shows one such case, where the slice reaches a libpcap function with a NULL variable even though the full program does not invoke the function under that precondition. Future work could reduce this limitation by incorporating caller-side constraints or lightweight precondition inference during wrapper generation.

\noindent
\textbf{Differential comparison depends on comparable wrappers.}
\df improves differential comparison by replaying inputs across paired pre-commit and post-commit slices. However, when a commit changes a function signature or input structure, the generated wrappers may consume the same input bytes differently across versions. This can reduce the effectiveness of replay and lead to false positives, as observed in RQ4. A possible mitigation is to generate version-aware wrappers that preserve stable input encodings across signature changes.

\noindent
\textbf{Bug oracles are limited to sanitizer-detectable failures.}
\df currently relies on runtime sanitizers, including AddressSanitizer, LeakSanitizer, UndefinedBehaviorSanitizer, and MemorySanitizer, to detect memory-safety bugs. These oracles are effective for failures such as buffer overflows, memory leaks, uninitialized-memory uses, and undefined behavior, but they do not capture higher-level semantic regressions. Future work could extend \df with additional runtime checks or generated assertions to detect semantic changes that do not trigger sanitizer reports.

\section{Related Work}
\label{sec}

\noindent
\textbf{Program slicing.}
Program slicing has long been used to reduce the amount of code that must be inspected or analyzed by preserving statements relevant to a target computation~\cite{Tip1994Jul,binkley1996program}. Prior work has studied static, dynamic, forward, backward, and path-sensitive slicing, as well as the trade-offs between slice precision, scalability, and context sensitivity~\cite{korel1988dynamic,Binkley,Binkley1,jaffar2012path}. These techniques have been used to support debugging, testing, program understanding, and fault localization~\cite{harman1995using,lei2013backward}.

\df also relies on slicing, but its goal is different. Rather than constructing a slice for a known bug location or a static-analysis warning, \df constructs paired slices for functions modified by a commit. The pre-commit slice provides a behavioral baseline, while the post-commit slice represents the changed code. This pairing enables \df to use differences in sanitizer-observed behavior as evidence that a report is related to the target commit.

\noindent
\textbf{Fuzzing and directed fuzzing.}
Fuzzing has been widely used to find memory-safety bugs by repeatedly executing programs with generated or mutated inputs~\cite{Li2018Dec,bohme2020fuzzing,BibEntry2023Mar,afl++}. Directed fuzzing techniques guide exploration toward specific target locations, such as known vulnerable code, patched regions, or locations reported by other analyses~\cite{bohme2017directed,Chen2018Oct,Wustholz2019May,Huang2022May,272157,DAFL}. These techniques are effective when the target location is known and the full program can be executed with suitable inputs and harnesses.

\df addresses a different use case. It does not assume a known warning or vulnerability location. Instead, the input is a commit, and the task is to determine whether the change introduces a new memory-safety bug. Whole-program fuzzing can be expensive for this use case and may not reach the changed code within a short budget~\cite{Bohme2020Nov,gao2023beyond}. \df therefore fuzzes executable slices around modified functions and compares the resulting sanitizer reports across versions.

\noindent
\textbf{Slice-based fuzzing and isolated execution.}
Recent work has explored fuzzing smaller code fragments or slices to improve reachability and reduce the cost of dynamic analysis~\cite{Chen2022Nov,FuzzSlice,TPSlice}. Slice-based fuzzing can start execution closer to the code of interest and can expose failures that are difficult to reach through whole-program inputs. However, isolated execution also introduces challenges. Because slices relax full-program constraints, they may expose failures that are not realizable in the original program.

\df builds on the efficiency benefits of slice-based fuzzing but adds a differential interpretation layer. A sanitizer report from an isolated slice is not reported directly. Instead, \df compares the post-commit report with the corresponding pre-commit slice. Reports that already appear before the commit are filtered out, while reports that appear only after the commit are retained as candidate commit-induced bugs. This differential design helps reduce noise caused by pre-existing bugs and slice-execution artifacts.

\noindent
\textbf{Static-analysis warning validation.}
Several techniques attempt to validate static-analysis warnings by using dynamic analysis, symbolic execution, machine learning, or fuzzing to determine whether a reported warning can manifest as a real failure~\cite{csallner2005check,TPSlice,FuzzSlice,Koc2017Jun,Yoon2014Dec}. These approaches are valuable for reducing false positives from static analyzers, which remain a major barrier to adoption~\cite{Johnson2013May,Kang2022May,Aloraini2019Dec,vassallo2020developers}. However, they typically start from a known warning location and focus on deciding whether that warning is feasible.

\df is not a static-warning validator. It is static-tool-agnostic and uses the commit itself as the unit of analysis. The goal is not to confirm a pre-existing warning, but to identify sanitizer-observed failures that newly appear after a code change. This makes \df complementary to warning-validation techniques: static analyzers can point developers to suspicious code, while \df can directly check whether a commit introduces new runtime memory-safety failures.

\noindent
\textbf{Commit-level and regression-oriented analysis.}
Prior work has studied the characteristics of commits, code evolution, and the relationship between software changes and defects~\cite{alali2008s,hindle2008large,eyolfson2014correlations,canfora2011long}. Regression testing and continuous integration help developers check whether recent changes break existing behavior, but they depend on tests that exercise the changed code~\cite{duvall2007continuous,Alegroth2016Feb,Levin2017Sep}. In practice, writing or maintaining such tests can be costly, especially for low-level bugs.

\df brings regression-style comparison to slice-based fuzzing. Instead of requiring a whole-program regression test that reaches the modified code, \df constructs paired fuzzing targets for the modified function before and after the commit. The pre-commit slice acts as the baseline, and the post-commit slice is checked for newly appearing sanitizer reports. This allows \df to support commit-level verification without requiring manually written tests or whole-program fuzzing harnesses.

%%%
\section{Conclusion}
\label{sec}

In this paper, we introduce \df, a commit-aware differential slice-based fuzzing method for commit-level verification. \df identifies functions modified by a target commit, constructs paired executable slices from the pre-commit and post-commit versions, fuzzes both slices, and compares their sanitizer reports. By using the pre-commit slice as a behavioral baseline, \df reports only failures that newly appear after the commit, turning slice fuzzing from a general bug-finding technique into an \textit{on-demand} commit verification method.

We evaluate \df on 300 commits from three widely used C projects. Across these commits, slice fuzzing initially produces 518 sanitizer reports. Differential filtering reduces these reports to 7 candidate commit-induced bug reports, of which 5 are confirmed as real bugs and 2 are classified as false positives. Several confirmed bugs have been fixed by project developers after we reported them.
\df analyzes a commit in 32.4 minutes on average and achieves 75.36\% average coverage of modified functions, showing that the approach can provide practical feedback for code review.

These results show that differential slice-based fuzzing can make commit-level memory-safety verification more focused and practical. By comparing paired pre-commit and post-commit slices, \df reduces noisy sanitizer output to a small set of reports relevant to recent code changes, helping developers prioritize bugs that are likely introduced or exposed by a commit.
%%%
%%%
%%%
% \section{Conclusion}
% In this paper, we introduced a novel differential fuzzing framework specifically designed for commit-level verification. By generating and fuzzing multiple code slices across target and baseline commits, and systematically comparing their runtime behaviors, our approach enables precise identification of vulnerabilities introduced by code changes. This slice-level differential analysis combines the efficiency of slicing-based fuzzing with the discriminative power of differential testing, allowing our framework to detect commit-induced bugs without requiring whole-program fuzzing or manually crafted testcases.

% Our empirical evaluation on real-world open-source projects demonstrates that the proposed approach is both practical and effective. Across 300 commits, the framework successfully identifies five bugs introduced by commits, while maintaining a low false positive rate of just 2 reported bugs. It scales to large codebases and operates within 32 minutes per commit, making it well suited for modern development workflows. We further analyze the sources of false positives and report encouraging feedback from open-source developers, highlighting its strong potential for real-world adoption.

\bibliographystyle{IEEEtranS}
\bibliography{Bibliography}

@incollection{Kang2022May,
	author = {Kang, Hong Jin and Aw, Khai Loong and Lo, David},
	title = {{Detecting false alarms from automatic static analysis tools: how far are we?}},
	booktitle = {{ICSE '22: Proceedings of the 44th International Conference on Software Engineering}},
	pages = {698--709},
	year = {2022},
	month = may,
	isbn = {978-1-45039221-1},
	publisher = {Association for Computing Machinery},
	address = {New York, NY, USA},
	doi = {10.1145/3510003.3510214}
}

@inproceedings{bohme2017directed,
  title={Directed greybox fuzzing},
  author={B{\"o}hme, Marcel and Pham, Van-Thuan and Nguyen, Manh-Dung and Roychoudhury, Abhik},
  booktitle={Proceedings of the 2017 ACM SIGSAC conference on computer and communications security},
  pages={2329--2344},
  year={2017}
}

@inproceedings{FuzzSlice,
  title={Fuzzslice: Pruning false positives in static analysis warnings through function-level fuzzing},
  author={Murali, Aniruddhan and Mathews, Noble and Alfadel, Mahmoud and Nagappan, Meiyappan and Xu, Meng},
  booktitle={Proceedings of the 46th IEEE/ACM International Conference on Software Engineering},
  pages={1--13},
  year={2024}
}

@incollection{Nadeem2012Mar,
	author = {Nadeem, Muhammad and Williams, Byron J. and Allen, Edward B.},
	title = {{High false positive detection of security vulnerabilities: a case study}},
	booktitle = {{ACM-SE '12: Proceedings of the 50th Annual Southeast Regional Conference}},
	pages = {359--360},
	year = {2012},
	month = mar,
	isbn = {978-1-45031203-5},
	publisher = {Association for Computing Machinery},
	address = {New York, NY, USA},
	doi = {10.1145/2184512.2184604}
}

@incollection{Park2016May,
	author = {Park, Joonyoung and Lim, Inho and Ryu, Sukyoung},
	title = {{Battles with False Positives in Static Analysis of JavaScript Web Applications in the Wild}},
	booktitle = {{2016 IEEE/ACM 38th International Conference on Software Engineering Companion (ICSE-C)}},
	journal = {2016 IEEE/ACM 38th International Conference on Software Engineering Companion (ICSE-C)},
	pages = {61--70},
	year = {2016},
	month = may,
	publisher = {IEEE},
	url = {https://ieeexplore.ieee.org/document/7883289}
}

@article{Ayewah2008Aug,
	author = {Ayewah, Nathaniel and Pugh, William and Hovemeyer, David and Morgenthaler, J. David and Penix, John},
	title = {{Using Static Analysis to Find Bugs}},
	journal = {IEEE Software},
	volume = {25},
	number = {5},
	pages = {22--29},
	year = {2008},
	month = aug,
	publisher = {IEEE},
	doi = {10.1109/MS.2008.130}
}

@incollection{Aloraini2017Sep,
	author = {Aloraini, Bushra and Nagappan, Meiyappan},
	title = {{Evaluating State-of-the-Art Free and Open Source Static Analysis Tools Against Buffer Errors in Android Apps}},
	booktitle = {{2017 IEEE International Conference on Software Maintenance and Evolution (ICSME)}},
	journal = {2017 IEEE International Conference on Software Maintenance and Evolution (ICSME)},
	pages = {295--306},
	year = {2017},
	month = sep,
	publisher = {IEEE},
	doi = {10.1109/ICSME.2017.77}
}

@incollection{Bohme2020Nov,
	author = {B{\ifmmode\ddot{o}\else\"{o}\fi}hme, Marcel and Falk, Brandon},
	title = {{Fuzzing: on the exponential cost of vulnerability discovery}},
	booktitle = {{ESEC/FSE 2020: Proceedings of the 28th ACM Joint Meeting on European Software Engineering Conference and Symposium on the Foundations of Software Engineering}},
	pages = {713--724},
	year = {2020},
	month = nov,
	isbn = {978-1-45037043-1},
	publisher = {Association for Computing Machinery},
	address = {New York, NY, USA},
	doi = {10.1145/3368089.3409729}
}

@article{serebryany2012addresssanitizer,
  title={Addresssanitizer: A fast address sanity checker},
  author={Serebryany, Konstantin and Bruening, Derek and Potapenko, Alexander and Vyukov, Dmitry},
  year={2012}
}

@article{koru2008investigation,
  title={An investigation into the functional form of the size-defect relationship for software modules},
  author={Koru, A G{\"u}ne{\c{s}} and Zhang, Dongsong and El Emam, Khaled and Liu, Hongfang},
  journal={IEEE Transactions on Software Engineering},
  volume={35},
  number={2},
  pages={293--304},
  year={2008},
  publisher={IEEE}
}

@inproceedings{edmundson2013empirical,
  title={An empirical study on the effectiveness of security code review},
  author={Edmundson, Anne and Holtkamp, Brian and Rivera, Emanuel and Finifter, Matthew and Mettler, Adrian and Wagner, David},
  booktitle={International Symposium on Engineering Secure Software and Systems},
  pages={197--212},
  year={2013},
  organization={Springer}
}

@inproceedings{alali2008s,
  title={What's a typical commit? A characterization of open source software repositories},
  author={Alali, Abdulkareem and Kagdi, Huzefa and Maletic, Jonathan I},
  booktitle={2008 16th IEEE international conference on program comprehension},
  pages={182--191},
  year={2008},
  organization={IEEE}
}

@inproceedings{hindle2008large,
  title={What do large commits tell us? a taxonomical study of large commits},
  author={Hindle, Abram and German, Daniel M and Holt, Ric},
  booktitle={Proceedings of the 2008 international working conference on Mining software repositories},
  pages={99--108},
  year={2008}
}

@inproceedings{godfrey2002tracking,
  title={Tracking structural evolution using origin analysis},
  author={Godfrey, Michael and Tu, Qiang},
  booktitle={Proceedings of the international workshop on Principles of software evolution},
  pages={117--119},
  year={2002}
}

@book{duvall2007continuous,
  title={Continuous integration: improving software quality and reducing risk},
  author={Duvall, Paul M and Matyas, Steve and Glover, Andrew},
  year={2007},
  publisher={Pearson Education}
}

@article{marinescu2012assessing,
  title={Assessing technical debt by identifying design flaws in software systems},
  author={Marinescu, Radu},
  journal={IBM Journal of Research and Development},
  volume={56},
  number={5},
  pages={9--1},
  year={2012},
  publisher={IBM}
}

@inproceedings{canfora2011long,
  title={How long does a bug survive? an empirical study},
  author={Canfora, Gerardo and Ceccarelli, Michele and Cerulo, Luigi and Di Penta, Massimiliano},
  booktitle={2011 18th Working Conference on Reverse Engineering},
  pages={191--200},
  year={2011},
  organization={IEEE}
}

@article{Alegroth2016Feb,
	author = {Al{\ifmmode\acute{e}\else\'{e}\fi}groth, Emil and Feldt, Robert and Kolstr{\ifmmode\ddot{o}\else\"{o}\fi}m, Pirjo},
	title = {{Maintenance of Automated Test Suites in Industry: An Empirical study on Visual GUI Testing}},
	journal = {arXiv},
	year = {2016},
	month = feb,
	eprint = {1602.01226},
	doi = {10.48550/arXiv.1602.01226}
}

@article{Levin2017Sep,
	author = {Levin, Stanislav and Yehudai, Amiram},
	title = {{The Co-Evolution of Test Maintenance and Code Maintenance through the lens of Fine-Grained Semantic Changes}},
	journal = {arXiv},
	year = {2017},
	month = sep,
	eprint = {1709.09029},
	doi = {10.48550/arXiv.1709.09029}
}

@misc{file-cve-2017-1000249,
	title = {{NVD - CVE-2017-1000249}},
	year = {2026},
	month = jun,
	note = {[Online; accessed 25. Jun. 2026]},
	url = {https://nvd.nist.gov/vuln/detail/CVE-2017-1000249}
}

@incollection{Chen2022Nov,
	author = {Chen, Libo and Cai, Quanpu and Ma, Zhenbang and Wang, Yanhao and Hu, Hong and Shen, Minghang and Liu, Yue and Guo, Shanqing and Duan, Haixin and Jiang, Kaida and Xue, Zhi},
	title = {{SFuzz: Slice-based Fuzzing for Real-Time Operating Systems}},
	booktitle = {{CCS '22: Proceedings of the 2022 ACM SIGSAC Conference on Computer and Communications Security}},
	pages = {485--498},
	year = {2022},
	month = nov,
	isbn = {978-1-45039450-5},
	publisher = {Association for Computing Machinery},
	address = {New York, NY, USA},
	doi = {10.1145/3548606.3559367}
}

@incollection{Yoon2014Dec,
	author = {Yoon, Jongwon and Jin, Minsik and Jung, Yungbum},
	title = {{Reducing False Alarms from an Industrial-Strength Static Analyzer by SVM}},
	booktitle = {{APSEC '14: Proceedings of the 2014 21st Asia-Pacific Software Engineering Conference - Volume 02}},
	pages = {3--6},
	year = {2014},
	month = dec,
	isbn = {978-1-47997426-9},
	publisher = {IEEE Computer Society},
	address = {USA},
	doi = {10.1109/APSEC.2014.81}
}

@incollection{Koc2017Jun,
	author = {Koc, Ugur and Saadatpanah, Parsa and Foster, Jeffrey S. and Porter, Adam A.},
	title = {{Learning a classifier for false positive error reports emitted by static code analysis tools}},
	booktitle = {{MAPL 2017: Proceedings of the 1st ACM SIGPLAN International Workshop on Machine Learning and Programming Languages}},
	pages = {35--42},
	year = {2017},
	month = jun,
	isbn = {978-1-45035071-6},
	publisher = {Association for Computing Machinery},
	address = {New York, NY, USA},
	doi = {10.1145/3088525.3088675}
}

@article{Aloraini2019Dec,
	author = {Aloraini, Bushra and Nagappan, Meiyappan and German, Daniel M. and Hayashi, Shinpei and Higo, Yoshiki},
	title = {{An empirical study of security warnings from static application security testing tools}},
	journal = {Journal of Systems and Software},
	volume = {158},
	pages = {110427},
	year = {2019},
	month = dec,
	issn = {0164-1212},
	publisher = {Elsevier},
	doi = {10.1016/j.jss.2019.110427}
}

@inproceedings{gao2023beyond,
  title={Beyond the coverage plateau: A comprehensive study of fuzz blockers (registered report)},
  author={Gao, Wentao and Pham, Van-Thuan and Liu, Dongge and Chang, Oliver and Murray, Toby and Rubinstein, Benjamin IP},
  booktitle={Proceedings of the 2nd International Fuzzing Workshop},
  pages={47--55},
  year={2023}
}

@inproceedings{afl++,
  title={$\{$AFL++$\}$: Combining incremental steps of fuzzing research},
  author={Fioraldi, Andrea and Maier, Dominik and Ei{\ss}feldt, Heiko and Heuse, Marc},
  booktitle={14th USENIX workshop on offensive technologies (WOOT 20)},
  year={2020}
}

@misc{BibEntry2023Mar,
	title = {{american fuzzy lop}},
	year = {2023},
	month = mar,
	note = {[Online; accessed 10. Mar. 2023]},
	url = {https://lcamtuf.coredump.cx/afl}
}

@incollection{Johnson2013May,
	author = {Johnson, Brittany and Song, Yoonki and Murphy-Hill, Emerson and Bowdidge, Robert},
	title = {{Why don't software developers use static analysis tools to find bugs?}},
	booktitle = {{2013 35th International Conference on Software Engineering (ICSE)}},
	journal = {2013 35th International Conference on Software Engineering (ICSE)},
	pages = {672--681},
	year = {2013},
	month = may,
	issn = {1558-1225},
	publisher = {IEEE},
	doi = {10.1109/ICSE.2013.6606613}
}

@incollection{Chen2018Oct,
	author = {Chen, Hongxu and Xue, Yinxing and Li, Yuekang and Chen, Bihuan and Xie, Xiaofei and Wu, Xiuheng and Liu, Yang},
	title = {{Hawkeye: Towards a Desired Directed Grey-box Fuzzer}},
	booktitle = {{CCS '18: Proceedings of the 2018 ACM SIGSAC Conference on Computer and Communications Security}},
	pages = {2095--2108},
	year = {2018},
	month = oct,
	isbn = {978-1-45035693-0},
	publisher = {Association for Computing Machinery},
	address = {New York, NY, USA},
	doi = {10.1145/3243734.3243849}
}

@article{Wustholz2019May,
	author = {W{\ifmmode\ddot{u}\else\"{u}\fi}stholz, Valentin and Christakis, Maria},
	title = {{Targeted Greybox Fuzzing with Static Lookahead Analysis}},
	journal = {arXiv},
	year = {2019},
	month = may,
	eprint = {1905.07147},
	doi = {10.48550/arXiv.1905.07147}
}

@incollection{Huang2022May,
	author = {Huang, Heqing and Guo, Yiyuan and Shi, Qingkai and Yao, Peisen and Wu, Rongxin and Zhang, Charles},
	title = {{BEACON: Directed Grey-Box Fuzzing with Provable Path Pruning}},
	booktitle = {{2022 IEEE Symposium on Security and Privacy (SP)}},
	journal = {2022 IEEE Symposium on Security and Privacy (SP)},
	pages = {36--50},
	year = {2022},
	month = may,
	issn = {2375-1207},
	publisher = {IEEE},
	doi = {10.1109/SP46214.2022.9833751}
}

@inproceedings {272157,
author = {Gwangmu Lee and Woochul Shim and Byoungyoung Lee},
title = {Constraint-guided Directed Greybox Fuzzing},
booktitle = {30th USENIX Security Symposium (USENIX Security 21)},
year = {2021},
isbn = {978-1-939133-24-3},
pages = {3559--3576},
url = {https://www.usenix.org/conference/usenixsecurity21/presentation/lee-gwangmu},
publisher = {USENIX Association},
month = aug,
}

@incollection{TPSlice,
	author = {Kallingal Joshy, Ashwin and Chen, Xueyuan and Steenhoek, Benjamin and Le, Wei},
	title = {{Validating static warnings via testing code fragments}},
	booktitle = {{ISSTA 2021: Proceedings of the 30th ACM SIGSOFT International Symposium on Software Testing and Analysis}},
	pages = {540--552},
	year = {2021},
	month = jul,
	isbn = {978-1-45038459-9},
	publisher = {Association for Computing Machinery},
	address = {New York, NY, USA},
	doi = {10.1145/3460319.3464832}
}

@article{eyolfson2014correlations,
  title={Correlations between bugginess and time-based commit characteristics},
  author={Eyolfson, Jon and Tan, Lin and Lam, Patrick},
  journal={Empirical Software Engineering},
  volume={19},
  pages={1009--1039},
  year={2014},
  publisher={Springer}
}

@article{harman1995using,
  title={Using program slicing to simplify testing},
  author={Harman, Mark and Danicic, Sebastian},
  journal={Software Testing, Verification and Reliability},
  volume={5},
  number={3},
  pages={143--162},
  year={1995},
  publisher={Wiley Online Library}
}

@article{louridas2006static,
  title={Static code analysis},
  author={Louridas, Panagiotis},
  journal={Ieee Software},
  volume={23},
  number={4},
  pages={58--61},
  year={2006},
  publisher={IEEE}
}

@article{korel1988dynamic,
  title={Dynamic program slicing},
  author={Korel, Bogdan and Laski, Janusz},
  journal={Information processing letters},
  volume={29},
  number={3},
  pages={155--163},
  year={1988},
  publisher={Elsevier}
}

@inproceedings{jaffar2012path,
  title={Path-sensitive backward slicing},
  author={Jaffar, Joxan and Murali, Vijayaraghavan and Navas, Jorge A and Santosa, Andrew E},
  booktitle={Static Analysis: 19th International Symposium, SAS 2012, Deauville, France, September 11-13, 2012. Proceedings 19},
  pages={231--247},
  year={2012},
  organization={Springer}
}

@misc{MSAN,
	title        = {{MemorySanitizer {\ifmmode---\else\textemdash\fi} Clang 23.0.0git documentation}},
	year         = 2026,
	month        = jan,
	url          = {https://clang.llvm.org/docs/MemorySanitizer.html},
	note         = {[Online; accessed 13. Jan. 2026]}
}

@misc{libpcap-fp-oob-commit,
	title = {{Lose the no-op dir qualifier in gen{$\_$}gateway() and gen{$\_$}proto(). {$\cdot$} the-tcpdump-group/libpcap@9823260}},
	year = {2026},
	month = mar,
	note = {[Online; accessed 29. Mar. 2026]},
	url = {https://github.com/the-tcpdump-group/libpcap/commit/98232604d3a3e79e91797fb53eda5d6d12a3fcf2}
}

@misc{libpcap-buffer-overread-commit,
	title = {{Generate cleaner code for unqualified "port" and "portrange". {$\cdot$} the-tcpdump-group/libpcap@119b248}},
	journal = {GitHub},
	year = {2026},
	month = mar,
	note = {[Online; accessed 28. Mar. 2026]},
	url = {https://github.com/the-tcpdump-group/libpcap/commit/119b2481cb364fea3b02c5ca90da480d4970e072}
}

@misc{leaksanitizer,
	title = {{LeakSanitizer {\ifmmode---\else\textemdash\fi} Clang 21.0.0git documentation}},
	year = {2025},
	month = jan,
	note = {[Online; accessed 27. May 2025]},
	url = {https://clang.llvm.org/docs/LeakSanitizer.html}
}

@inproceedings{DAFL,
  title={$\{$DAFL$\}$: Directed Grey-box Fuzzing guided by Data Dependency},
  author={Kim, Tae Eun and Choi, Jaeseung and Heo, Kihong and Cha, Sang Kil},
  booktitle={32nd USENIX Security Symposium (USENIX Security 23)},
  pages={4931--4948},
  year={2023}
}

@misc{treesitter,
	title = {{Introduction - Tree-sitter}},
	year = {2025},
	month = apr,
	note = {[Online; accessed 27. May 2025]},
	url = {https://tree-sitter.github.io/tree-sitter}
}

@article{binkley1996program,
  title={Program slicing},
  author={Binkley, David W and Gallagher, Keith Brian},
  journal={Advances in computers},
  volume={43},
  pages={1--50},
  year={1996},
  publisher={Elsevier}
}

@inproceedings{lei2013backward,
  title={Backward-slice-based statistical fault localization without test oracles},
  author={Lei, Yan and Mao, Xiaoguang and Chen, Tsong Yueh},
  booktitle={2013 13th International Conference on Quality Software},
  pages={212--221},
  year={2013},
  organization={IEEE}
}

@incollection{Binkley,
	author = {Binkley, D. and Harman, M.},
	title = {{Forward slices are smaller than backward slices}},
	booktitle = {{Fifth IEEE International Workshop on Source Code Analysis and Manipulation (SCAM'05)}},
	pages = {2005--01},
	isbn = {978-0-7695-2292},
	publisher = {IEEE},
	doi = {10.1109/SCAM.2005.28}
}

@incollection{Binkley1,
	author = {Binkley, D. and Harman, M.},
	title = {{A large-scale empirical study of forward and backward static slice size and context sensitivity}},
	booktitle = {{International Conference on Software Maintenance, 2003. ICSM 2003. Proceedings.}},
	pages = {22--26},
	isbn = {978-0-7695-1905},
	publisher = {IEEE},
	doi = {10.1109/ICSM.2003.1235405}
}

@article{bohme2020fuzzing,
  title={Fuzzing: Challenges and reflections},
  author={B{\"o}hme, Marcel and Cadar, Cristian and Roychoudhury, Abhik},
  journal={IEEE Software},
  volume={38},
  number={3},
  pages={79--86},
  year={2020},
  publisher={IEEE}
}

@article{vassallo2020developers,
  title={How developers engage with static analysis tools in different contexts},
  author={Vassallo, Carmine and Panichella, Sebastiano and Palomba, Fabio and Proksch, Sebastian and Gall, Harald C and Zaidman, Andy},
  journal={Empirical Software Engineering},
  volume={25},
  pages={1419--1457},
  year={2020},
  publisher={Springer}
}

@article{zheng2006value,
  title={On the value of static analysis for fault detection in software},
  author={Zheng, Jiang and Williams, Laurie and Nagappan, Nachiappan and Snipes, Will and Hudepohl, John P and Vouk, Mladen A},
  journal={IEEE transactions on software engineering},
  volume={32},
  number={4},
  pages={240--253},
  year={2006},
  publisher={IEEE}
}

@inproceedings{csallner2005check,
  title={Check'n'crash: combining static checking and testing},
  author={Csallner, Christoph and Smaragdakis, Yannis},
  booktitle={Proceedings of the 27th international conference on Software engineering},
  pages={422--431},
  year={2005}
}

@article{Li2018Dec,
	author = {Li, Jun and Zhao, Bodong and Zhang, Chao},
	title = {{Fuzzing: a survey}},
	journal = {Cybersecur.},
	volume = {1},
	number = {1},
	pages = {1--13},
	year = {2018},
	month = dec,
	issn = {2523-3246},
	publisher = {SpringerOpen},
	doi = {10.1186/s42400-018-0002-y}
}

@book{Tip1994Jul,
	author = {Tip, Frank},
	title = {{A Survey of Program Slicing Techniques.}},
	journal = {Guide books},
	year = {1994},
	month = jul,
	publisher = {CWI (Centre for Mathematics and Computer Science)},
	doi = {10.5555/869354}
}

@misc{UBSAN,
	title = {{UndefinedBehaviorSanitizer {\ifmmode---\else\textemdash\fi} Clang 17.0.0git documentation}},
	year = {2023},
	month = jun,
	note = {[Online; accessed 27. Jun. 2023]},
	url = {https://clang.llvm.org/docs/UndefinedBehaviorSanitizer.html}
}

@misc{openssl-buffer-overflow-commit,
	title = {{ossl{$\_$}property{$\_$}list{$\_$}to{$\_$}string: handle quoted strings {$\cdot$} vdukhovni/openssl@fb20e66}},
	journal = {GitHub},
	year = {2026},
	month = mar,
	note = {[Online; accessed 23. Mar. 2026]},
	url = {https://github.com/vdukhovni/openssl/commit/fb20e66c6b2651067f50bab8cf098c71e2caed4b}
}

@misc{leptonica-mem-leak-commit,
	title = {{Resolve Issue {\#}734: invalid colormap made in jbcorrelation and jbrank{\ldots} {$\cdot$} DanBloomberg/leptonica@5e4f9a6}},
	year = {2026},
	month = mar,
	note = {[Online; accessed 23. Mar. 2026]},
	url = {https://github.com/DanBloomberg/leptonica/commit/5e4f9a6dba8a440689edb803b19d72648282e270}
}

@misc{leptonica-integer-overflow-commit,
	title = {{Fix a few errors found by Coverity Scan {$\cdot$} DanBloomberg/leptonica@def39f1}},
	journal = {GitHub},
	year = {2026},
	month = mar,
	note = {[Online; accessed 22. Mar. 2026]},
	url = {https://github.com/DanBloomberg/leptonica/commit/def39f10cea8bf3fcfe796ee40ed28902a318b09}
}

\end{document}